\documentstyle[12pt,dina4,psfig,times]{article}
\begin{document}
\parindent0.25cm
\renewcommand{\arraystretch}{1.4}
\newcommand{\be}{\begin{equation}}
\newcommand{\ee}{\end{equation}}
\newcommand{\ba}{\begin{array}}
\newcommand{\ea}{\end{array}}
\newcommand{\bea}{\begin{eqnarray}}
\newcommand{\eea}{\end{eqnarray} \nonumber}
\newcommand{\PP}{\rule{0.29mm}{2.65mm} \rm P }
\newcommand{\sPP}{\rule{0.5mm}{0mm} \sss \rule{0.25mm}{1.9mm} \rm P }
\newcommand{\PPS}{\rule{0.29mm}{2.65mm} \rm P \!\!\!\! / \rule{1.1mm}{0mm}}
\newcommand{\KP}{\rm P}
\newcommand{\KPS}{\rm P \hspace{-2.3mm} / \rule{1.3mm}{0mm}}
\newcommand{\MM}{\rm M_{\Phi}}
\newcommand{\JP}{J/\Psi}
\newcommand{\LS}{l \hspace{-1.5mm} / \rule{0.05mm}{0mm}}
\newcommand{\LSS}{l^{\,\prime} \hspace{-2.9mm} / \rule{0.9mm}{0mm}}
\newcommand{\ps}{p \hspace{-1.7mm}  / \rule{0.3mm}{0mm}}
\newcommand{\pss}{p{\prime} \hspace{-2.85mm} / \rule{1.1mm}{0mm}}
\newcommand{\ds}{\displaystyle}
\newcommand{\sss}{\scriptstyle}
\newcommand{\OO}{\rule{0.75mm}{0mm} \rule{0.29mm}{3.1mm} \!\!
                 \rule{0.35mm}{0.0mm}  \rm O }
\newcommand{\sOO}{\rule{0.4mm}{0mm} \sss \rule{0.25mm}{2.1mm} \!\!\!
                  \rule{0.5mm}{0mm} \rm O \rule{0.3mm}{0mm} }
\begin{center}
{\huge Diffractive Meson Production\\[0.5cm]
and the  \\[0.5cm]
Quark-Pomeron Coupling}
\\[2cm]
{\Large J.~Klenner, A.~Sch\"afer, W.~Greiner}
\\[0.4cm]
{\it Institut f\"ur Theoretische Physik, Universit\"at Frankfurt am Main,
Postfach 111 932,\\
D-60054 Frankfurt am Main, Germany}
\end{center}
\vspace{3cm}
\begin{center}
   {\bf Abstract}
\end{center}

$\vspace{0.2cm}$

Diffractive meson production at HERA offers
interesting possibilities to
investigate diffractive processes and thus to learn
something about the properties of the pomeron.
The most succesful phenomenological description of the
pomeron so far assumes it to couple like a
$C = +1$ isoscalar photon to single quarks.
This coupling leads, however, to problems for exclusive
diffractive reactions. We propose a new phenomenological pomeron vertex,
which leads to very good fits to the known data,
but avoids the problems of the old vertex.

\newpage

\section{Introduction}

The start of HERA has opened very promising possibilities to
investigate the nature of the pomeron \cite{hera}. For example the
new data for $F_{2}$ seem to
favour the
BFKL--pomeron which is calculated within perturbative QCD,
but the Donnachie--Landshoff pomeron is not ruled out
definitely \cite{f2}.
The latter, however, gives a perfect description of all aspects of
diffractive hadronic reactions, which the BFKL--pomeron does not.
In addition the BFKL--pomeron is problematic for theoretical
reasons \cite{hans}.

We do not want to investigate the relation between the pomeron
and $F_{2}$ or the consequences of the different pictures
for the structure functions. Instead we
investigate a certain class of processes in some detail, namely
diffractive meson production in lepton-nucleon and
nucleon-nucleon scattering.
We start from the Donnachie and Landshoff (DL) pomeron,
and concentrate on the coupling of
the pomeron to single quarks.

The diffractive $\JP$ production in nucleon-nucleon collisions
was already suggested \cite{jpsi1} as an attractive possibility to learn
more about the pomeron's coupling to single quarks in
hadronic matter. The background from $\chi$ production
has been calculated too \cite{ecki}.

We have tried to apply a similar program for diffractive events at HERA,
but it turned out that the traditonal DL--pomeron cannot describe
them in an unambigous way.

The failure of the classical pomeron description
motivated us to investigate
its coupling to quarks, suggesting a new effective vertex.
This new vertex is used to calculate several cross sections for diffractive
production of vector mesons, in lepton--nucleon
as well as in nucleon--nucleon scattering. The results offer definite
possibilities to distinguish between the traditional picture and
the new one.

In addition we guarantee that well established results as total
nucleon--nucleon cross sections and elastic proton--proton
cross sections are described well by the modified coupling.

The last process we discuss is the exclusive $\rho$ production in DIS.
For this process the situation is rather unclear.
The results for both vertices look qualitively fine, but the
normalization is somewhat off.
This point needs further clarification.
It is certainly related to the fact that here the quark masses
are much smaller than $M_{\rho} / 2$.
The fraction of longitudinally polarised
$\rho$ mesons can be described in both pictures.

In summary the modified coupling allows to describe $\Phi$ production
at HERA in a reasonable way and it predicts clear signals
for $\JP$ production in
nucleon--nucleon collisions. Therefore it would be of great interest to get
experimental data for these processes.

\section{Diffractive meson production at HERA}

\begin{figure}[p]
   \par
   \centerline{\psfig{figure=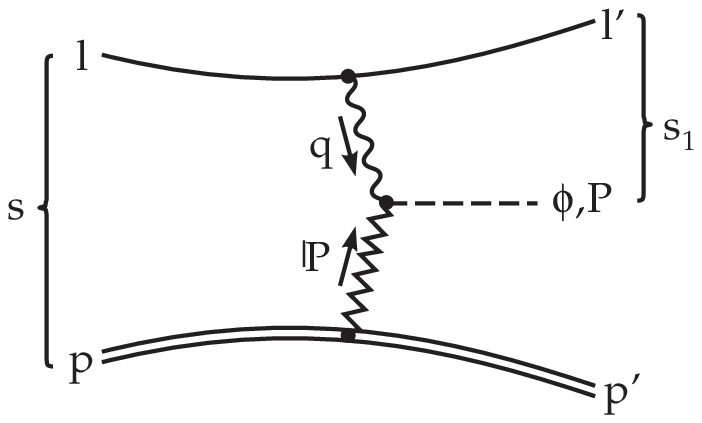,width=9cm}}
   \par
   \caption{Diffractive $\Phi$ production at HERA}
  \label{phi}
\end{figure}

The production of $J^{PC}=1^{--}$ mesons in diffractive events at HERA
opens
new possibilities to test the coupling of the pomeron to single
quarks. Figure \ref{phi} shows the relevant Feynman graph,
with the four momenta attributed to the involved particles.
In the following
we consider the production of $\Phi(1020)$ mesons. For $J/\Psi$ the
mechanism is the same, but the cross sections are smaller due to the
larger mass and the smaller size of this meson.

\begin{figure}[p]
   \par
   \centerline{\psfig{figure=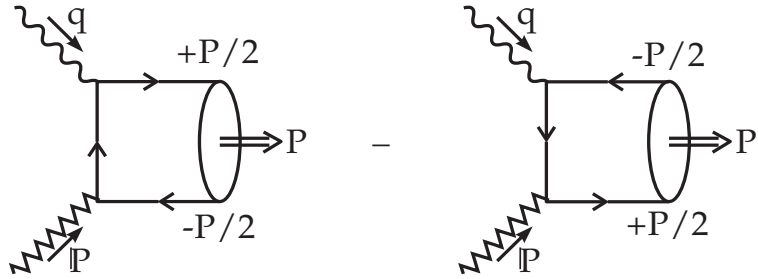,width=10cm}}
   \par
   \caption{The two contributions to the
            $\gamma $-$\protect{\PP} $-$\Phi $-vertex}
  \label{phivert}
\end{figure}

The $\Phi$ meson is described as a non relativistic bound state of a
$s\overline{s}$ pair, figure \ref{phivert}.
There are two contributions for the different
directions of momentum flow, and since the pomeron has $C=+1$ the two
amplitudes have to be subtracted. A formal derivation of this sign
is given in \cite{jpsi1}. Note that for two--photon fusion both
amplitudes would have to be added coherently and the sum would vanish
identically \cite{kuhn}.

We use the normalization of \cite{berger,jpsi1} and take the following
expression for the $q$-$\PP$-$\Phi$ vertex:
\bea
 \label{tmak}
  T_{\mu\alpha} & = & e_{q}\,\beta \, \sqrt{3} A \\
 & \times & {\rm\bf tr}
    \left\{ \varepsilon_{\Phi} \!\!\!\!\!\! / \,\,\, (\KPS + \MM)
     \left[\gamma_{\mu}\frac{-\PPS + \frac{1}{2}\KPS + m_{s}}
                           {(\frac{1}{2}\KP - \PP)^{2} - m_{s}^{2}}
           \Gamma_{\alpha}  -
      \Gamma_{\alpha}\frac{\PPS - \frac{1}{2}\KPS + m_{s}}
                           {(-\frac{1}{2}\KP + \PP)^{2} - m_{s}^{2}}
           \gamma_{\mu} \right] \right\} \; . \nonumber
\eea
The normalization constant $A$ can be related to the $e^{+}e^{-}$
decay width
\be
 \label{psi0}
  3 A^{2} \; = \; 3 \, \frac{\ds 2\pi  \Gamma_{ee} \MM}
                        {\ds e_{q}^{2} e^{2}} \, ,
\ee
where the factor $3$ counts the colours and $e_q$ is the electric charge
of the $s$-quark. The strength of the pomeron's coupling to a
single quark is described by the "charge" $\beta$. $\gamma_{\mu}$ is the
usual quark-photon vertex while $\Gamma_{\alpha}$ is the quark-pomeron
vertex, which we want to investigate in more detail.

\begin{figure}[p]
   \par
   \centerline{\psfig{figure=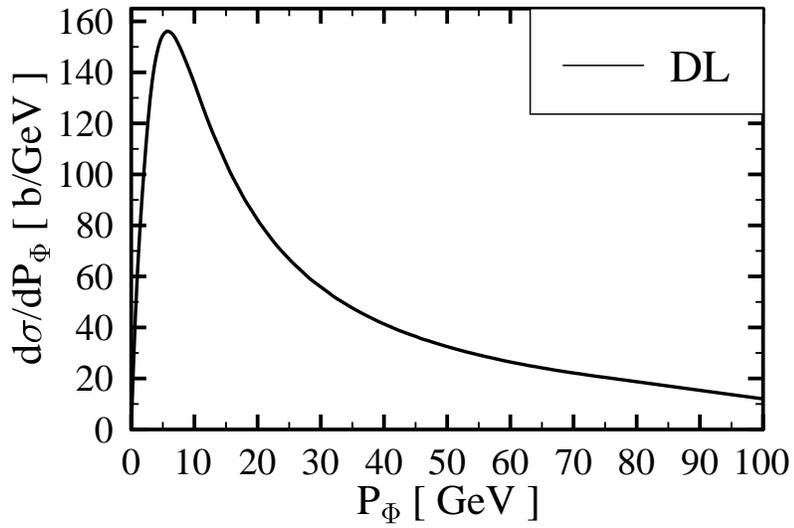}}
   \par
   \caption{The differential cross section
            for diffractive $\Phi$ production at HERA,
            calculated with the $\gamma$-coupling}
  \label{sig1}
\end{figure}

With these conventions the squared Feynman amplitude  reads
\be
  \left| {\cal M} \right|^{2} \; = \;
      \sum\limits_{\varepsilon_{\Phi}}
      \left[ \frac{\ds 1}{\ds q^{2}} \left(\frac{\ds s}{\ds s_{1}}
                \right)^{\alpha_{\sPP} - 1}
                F_{N}\left( \PP^{2}\right) \right]^{2}
       \, l^{\mu\nu} h^{\alpha\beta}
          T_{\mu\alpha} T^{*}_{\nu\beta} \, ,
\ee
with the usual leptonic tensor
\be
   l^{\mu\nu} \; = \; \frac{1}{2} \, {\rm \bf tr}
       \left[ (\LS + m_{e}) \gamma^{\mu}
               (\LSS + m_{e}) \gamma^{\nu} \right]
\ee
and the hadronic tensor
\be
   h^{\alpha\beta} \; = \; \frac{1}{2} \,
              {\rm \bf tr} \left[ (\ps + m_{N}) \Gamma^{\alpha}
                              (\pss + m_{N}) \Gamma^{\beta}
                     \right] \, .
\ee
The quantity $s_{1}$, used to normalise $s$ in the pomeron
propagator, is defined by \cite{higgs}
\be
   s_{1} \; = \; (l^{\prime} + \KP)^{2} \, .
\ee
The pomeron trajectory is taken to be linear \cite{pommi,l1}
\be
   \alpha_{\sPP}\left( t_{\sPP}\right) \; = \;
     1 + \varepsilon + \alpha^{\prime} t_{\sPP}\, ,
\ee
with the following parameters
\be
   \varepsilon \; = \; 0.08 \, , \quad
       \alpha^{\prime} \; = \; 0.25 \, {\rm GeV}^{-2} \, ,
\ee
and the "charge" $\beta$ has the value
\be
   \beta \; = \; 1.8 \, {\rm GeV}^{-1} \, .
\ee
The isoscalar form factor at the proton-pomeron vertex
is written in the
dipole approximation as
\be
   F_{N}(t) \; = \;
      \frac{4 m_{N}^{2} - 2.8 t}
           {4 m_{N}^{2} - t}
       \left( 1 -
            \frac{t}{0.7 \,{\rm GeV}^{2}}
                                    \right)^{-2} \,  .
\ee
To evaluate the differential cross section we integrate over the
appropriate
phase space.
This is done by using a Monte Carlo algorithm.
All our results are calculated exactly in this sense without any
approximation.
For the DL--vertex, $\Gamma_{\alpha} = \gamma_{\alpha}$,
the shape of the resulting differential cross section,
shown in figure \ref{sig1}, has the expected form
but the absolute values are in the region of several barns, which can not
be true.

This problem can be traced back to the violation of gauge invariance.
In QCD the pomeron is in principle described by the exchange of two gluons and
we should take into account all six
permutations of the coupling of
two gluons and a photon to a quark-line.
The single contributions would cancel, and the resulting
cross sections would be of the order of several nb.
Since we use the effective description of the pomeron,
we do not take into account all graphs and the cancellation does not
take place.

To resolve this problem and to keep the
simple picture of the pomeron coupling to single quarks,
one could introduce a form factor for
the $\Phi$-vertex.
But a reasonable guess does not change the results dramatically.

We tried another possibility, namely a modification
of the effective quark-pomeron coupling. With the four momenta
$k$ and $k^{\prime} = k + \PP$ for incoming and outgoing quark
we introduce the new effective vertex, given by
\be
  \label{wawa}
    \Gamma_\alpha \; = \;
        {\ds\frac{1}{2}} \left(
      \gamma_\alpha +
        \frac{\mid k + k^\prime \pm \PP\mid }
             {2\left(\mid {\it k}^2\mid + \mid {\it k}^{\prime 2}\mid\right)}
         \left( k + k^\prime\right)_\alpha \right)
       \, , \quad
   \left| k + k^{\prime} \pm \PP \right| \; = \;
     \sqrt{ \left( k + k^{\prime} \pm \PP \right)^{2}} \, .
\ee
This definition implicitly makes use of the fact, that the
pomeron couples to single constituent quarks in hadrons,
corresponding to the well established quark counting rule. In this simple and
very successful picture the constituent quarks carry a nonvanishing mass.
E.g.~for mesons $m_q = M/2$ and for the nucleon we have $m_q = m_N/3$.
The two signs correspond to the coupling to quark and antiquark,
respectively. It turns out that
this choice of the coupling
is unique, i.e.~this is the only possibility to obtain
sensible results for diffractive meson production.
In most of the processes we studied small momentum transfers
dominate, and equation (\ref{wawa}) reduces to
a much simpler expression.

E.g.~for the two graphs in figure \ref{phivert}
the quark-pomeron vertices read
\be
 \label{jogi}
  \Gamma_{\alpha}^{(1)} \; = \;
       \frac{1}{2} \left( \gamma_{\alpha}
                           - \frac{{\rm M}_\Phi}
                      {{\rm M}_\Phi^{2} - q^{2} - \PP^{2}}
                                  q_{\alpha} \right)
   \quad \, {\rm and} \quad \,
  \Gamma_{\alpha}^{(2)} \; = \;
       \frac{1}{2} \left( \gamma_{\alpha}
                           + \frac{{\rm M}_\Phi}
                {{\rm M}_\Phi^{2} - q^{2} - \PP^{2}}
                          q_{\alpha} \right) \, ,
\ee
and for diffractive meson production only small values of
$-q^2,-\PP^2 \ll M^2_\Phi$ are involved, resulting
in the following effective form
\be
 \label{saml}
  \Gamma_{\alpha}^{(1)} \; \simeq \;
       \frac{1}{2} \left( \gamma_{\alpha}
                           - \frac{1}{\MM} q_{\alpha} \right)
   \quad \, {\rm and} \quad \,
  \Gamma_{\alpha}^{(2)} \; \simeq \;
       \frac{1}{2} \left( \gamma_{\alpha}
                           + \frac{1}{\MM} q_{\alpha} \right) \, .
\ee
The hadronic tensor has to be modified analogously by
substituting
\be
   \Gamma_{\alpha} \; \simeq \; \frac{1}{2}
      \left( \gamma_{\alpha} + \frac{1}{2 m_{N}}
                  \left( p + p^{\prime} \right)_{\alpha}
           \right) \, .
\ee
This nucleon pomeron vertex is derived by interpreting the nucleon
as consisting of three constituent quarks, each one carrying a third of
the nucleon's momentum. Consistently the quark mass is
taken to be a third of the nucleon's mass.

The corresponding changes for the hadronic tensor do
not lead to pronounced effects, for unpolarised reactions.
Nevertheless we
take them into account to be able to analyse polarization effects in
section 5.

\begin{figure}[p]
   \par
   \centerline{\psfig{figure=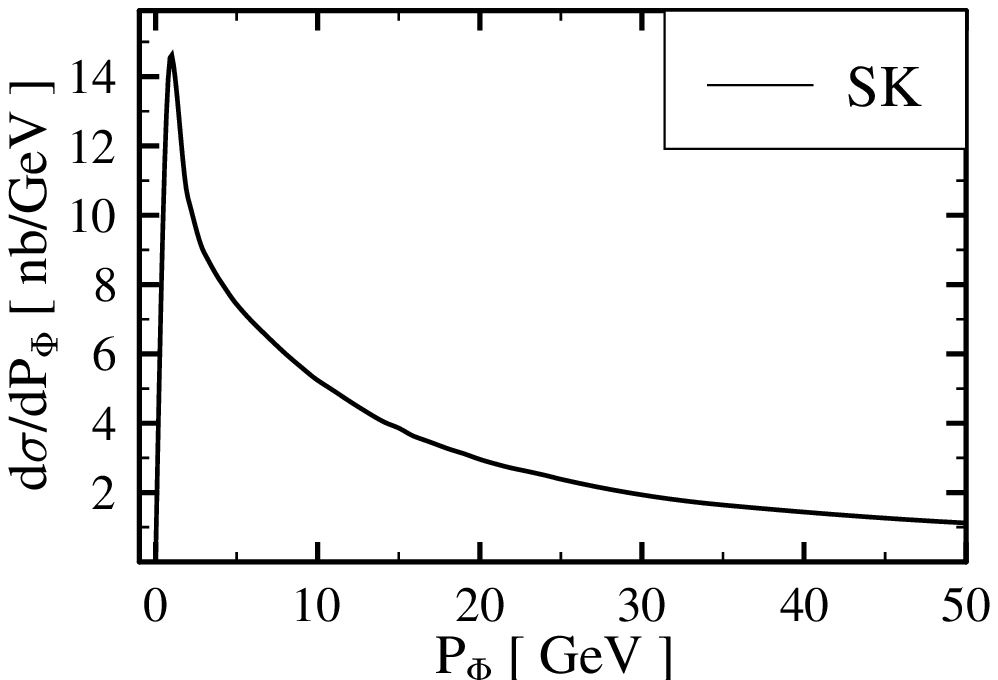}}
   \par
   \caption{The differential cross section
            for diffractive $\Phi$ production at HERA,
            calculated with the modified coupling}
  \label{sig2}
\end{figure}

Figure \ref{sig2} shows the differential cross section obtained with the
new coupling. Now the results look plausible, not only
the shape but the absolute values too.

This first success encourages us to suggest (\ref{wawa}) as
an alternative for the pomeron--quark coupling.
In the following we work out the consequences for
several processes.

First we should have a look at the vertex in some detail. The Gordon
identity relates the $\gamma$-coupling and the momentum
dependent terms:
\be
  \label{gordon}
  \overline{u} \, (k^{\prime}) \gamma_{\alpha}\, u(k)
     \; = \;
  \overline{u} (k^{\prime})
          \left[ \frac{1}{2 m_{N}} \left( k + k^{\prime} \right)_{\alpha}
                + \frac{i}{2 m_{N}} \sigma_{\alpha\alpha^{\prime}}
                      \PP^{\alpha^{\prime}} \right]
                u(k) \, ,
\ee
with the transferred momentum $\PP$. The two expressions differ by a term
containing the spin matrices $\sigma_{\alpha\alpha^{\prime}}$,
which could in principle generate spin dependent cross sections.

The first possibility we have is to fix the helicity of incoming and
outgoing electron and proton and to sum over all polarizations of the
produced $\Phi$. The helicity operator, e.g.~for the incoming electron,
is given by
\be
   \Sigma_{e} \; = \; \frac{1}{2}
      \left( 1 + \gamma_{5} s\!\!\! / \,\, \right)\, ,
    \quad
    s^{\mu} \; = \; (| \vec{l}\, |, E_{l} \, \hat{l} )
   \, ,
\ee
with energy $E_{l}$ and the unit vector $\hat l$ in the
direction of $\vec{ l}$ \cite{bjorken}.

Since $m_{e}/E_{l} \ll 1$ the spin flip of the electron is strongly
suppressed. This effect stems solely from the vertex at  the  electronic
current. Therefore we fix the helicity for the electron to remain
unchanged after the scattering.

\begin{figure}[p]
   \par
   \centerline{\psfig{figure=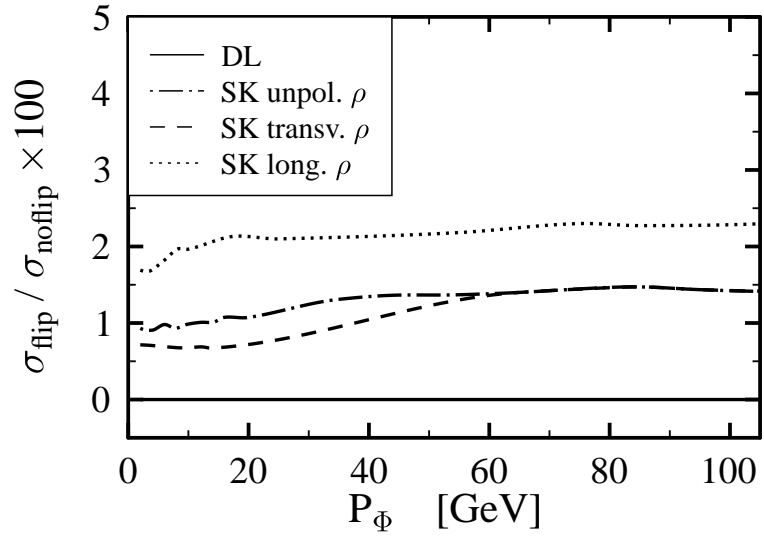}}
   \par
   \caption{Ratio of $\sigma_{\rm flip}$ and
            $\sigma_{\rm noflip}$, see text}
  \label{flip1}
\end{figure}

For a pure $\gamma$-coupling of the pomeron the flip of the proton's
helicity is suppressed strongly too, as the solid curve of figure
\ref{flip1} shows.
It represents the proportion of the differential cross sections for
helicity flip and no flip within the traditional description (DL).
The dashed-dotted curve shows the same quantity but calculated with the
modified vertex (SK). Here the helicity flip is enhanced to about
one percent
of the cross section for unchanged helicity.

A second interesting possibility is to look at the $\Phi$'s polarization
in some detail. For transverse
mesons $\sigma_{flip}/\sigma_{noflip}$ is about $0.7\%$ for small
momenta, raising to about $1.5\%$ at $P_{\Phi} = 100 $ GeV, while
this quantity is nearly constant at $2\%$ for a longitudinally polarised
$\Phi$, figure \ref{flip1}.

\begin{figure}[p]
   \par
   \centerline{\psfig{figure=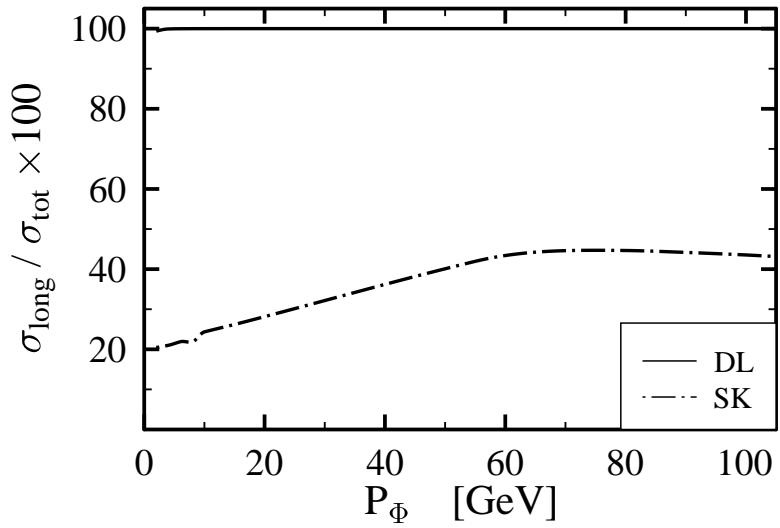}}
   \par
   \caption{Fraction of longitudinally polarised $\Phi$ mesons}
  \label{fraclong}
\end{figure}

Since the differences are small,
the experiments at HERA probably could not detect them.
But there is another interesting
quantitity leading to a clear signal.
In figure \ref{fraclong} we show the
fraction of longitudinally produced $\Phi$'s
in the
scattering of longitudinally polarised electrons and protons.
The solid curve for DL shows that nearly all mesons are produced in a
longitudinal state in the traditional picture.
For the new vertex
only a fraction of about $30\% \pm 10\%$ is
produced longitudinally.
Measuring the $\Phi$ polarization, e.g.~by analysing the angular
correlations of its decay products, if feasible, would
therefore be a very important experiment for pomeron
phenomenology.

\begin{figure}[p]
   \par
   \centerline{\psfig{figure=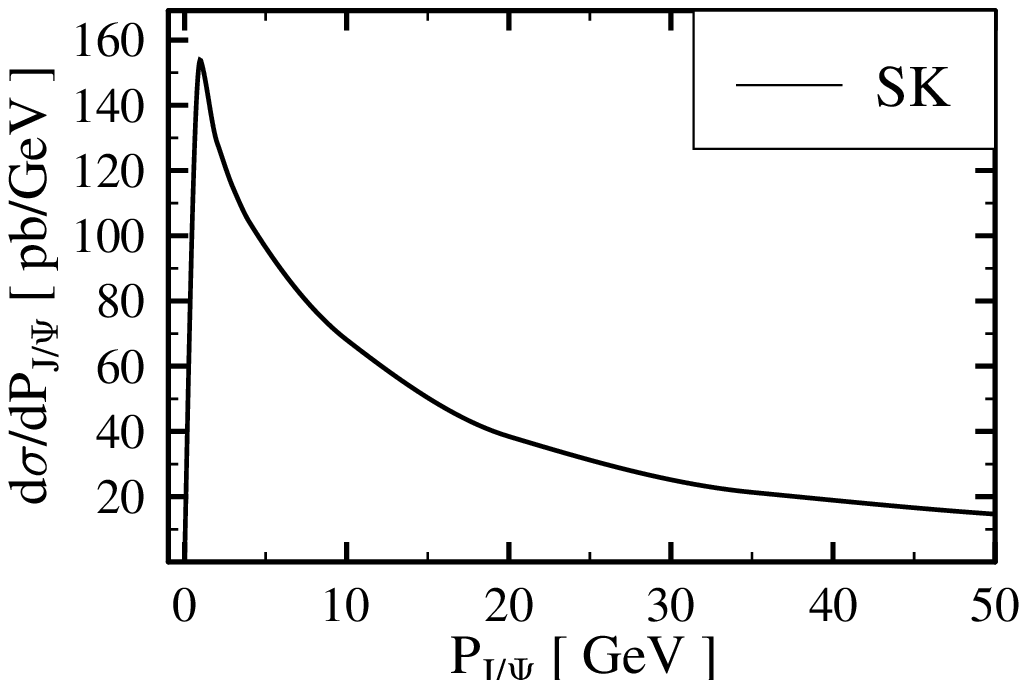}}
   \par
   \caption{The differential cross section
            for diffractive $\JP$ production at HERA,
            calculated with the modified coupling}
  \label{jpdiffo}
\end{figure}

For $\JP$ production at HERA the first experimental data
for the total cross section have just been
published \cite{jpex}, the result is $\sigma_{tot} = 8.8 \pm 2.0 \pm 2.2$ nb.
In figure \ref{jpdiffo} we show
the differential cross section for diffractive production,
calculated with the vertex (\ref{wawa}).
The results include an ad hoc factor
$N=0.1$ to take into account the small size
of the $\JP$ meson. This factor corresponds to the effect of some
unspecified pomeron form factor suppressing diffractive processes
when the produced meson is smaller than the pomeron.
Its
value is chosen very
small such that our result has
to be understood as a lower bound. The
resulting value for the integrated cross section is
$\sigma_{diff}= 4.0 \pm 1.0$ nb.
The error corresponds to the
cut--off for higher meson momenta where the
usual pomeron description is not applicable \cite{l1}.
Bearing in mind the uncertainty parametrised by the factor $N$
the order of magnitude of our result looks fine.

Finally we note an interesting property of the new vertex (\ref{wawa}).
The second term in parantheses
invalidates the usual relationship between diffractive scattering
and the behaviour of $F_{2}(x)$ for small $x$, depending on the
adapted $i\eta$--procedure \cite{parton}. This might open a loophole
if the $F_{2}(x)$--data were
found to
exclude definitely an
asymptotic behaviour  $x^{-\epsilon}$.

\section{Nucleon-nucleon cross sections}

The $\Gamma$-coupling introduced in the last section gives reasonable
results for diffractive production  of
$\Phi$ mesons at HERA, but
this is not the only possibility to test the new vertex, since
there exists a lot of experimental data described by
the DL-pomeron in outstanding agreement.
Maybe the most important example is the total cross
section for nucleon-nucleon scattering. In the traditional picture
the total cross section is described by \cite{l4}
\be
  \label{stot}
   \sigma_{\rm NN}^{tot} \; = \; A_{\rm NN}\,
     s^{-0.56}\, + \, B s^{\varepsilon} \, .
\ee
The first term describes $\rho,\omega,..$-exchange and the coefficient
$A_{\rm NN}$ is different for $pp$ and $p\bar{p}$ scattering.
The second term corresponding to pomeron exchange is the same for both cross
sections and is related to the elastic cross section through the
optical theorem:
\be
 \label{ot}
  \sigma^{tot} \; \sim \;
    \Im \left( \left. {\cal M}^{el} \right|_{t = 0} \right) \; = \;
       \Im \left( \left.
     (3 \beta)^{2} F^{2}_{N}(t) J_{\alpha} J^{\alpha}
       \left( \frac{s}{m_{N}^{2}} \right)^{\alpha_{\sPP} - 1}
       \right|_{\sPP = 0}
       \, + \quad \rho,\omega,... \right) \, ,
\ee
where $J_{\alpha}$  is the hadronic current. Since the amplitude
is evaluated for vanishing pomeron four momentum, $\PP=0$, the Gordon identity
(\ref{gordon}) ensures that the new vertex reproduces the
result (\ref{stot}). This calculation was used to fix the normalization
factor $\frac{1}{2}$ for the new vertex.

\begin{figure}[p]
   \par
   \centerline{\psfig{figure=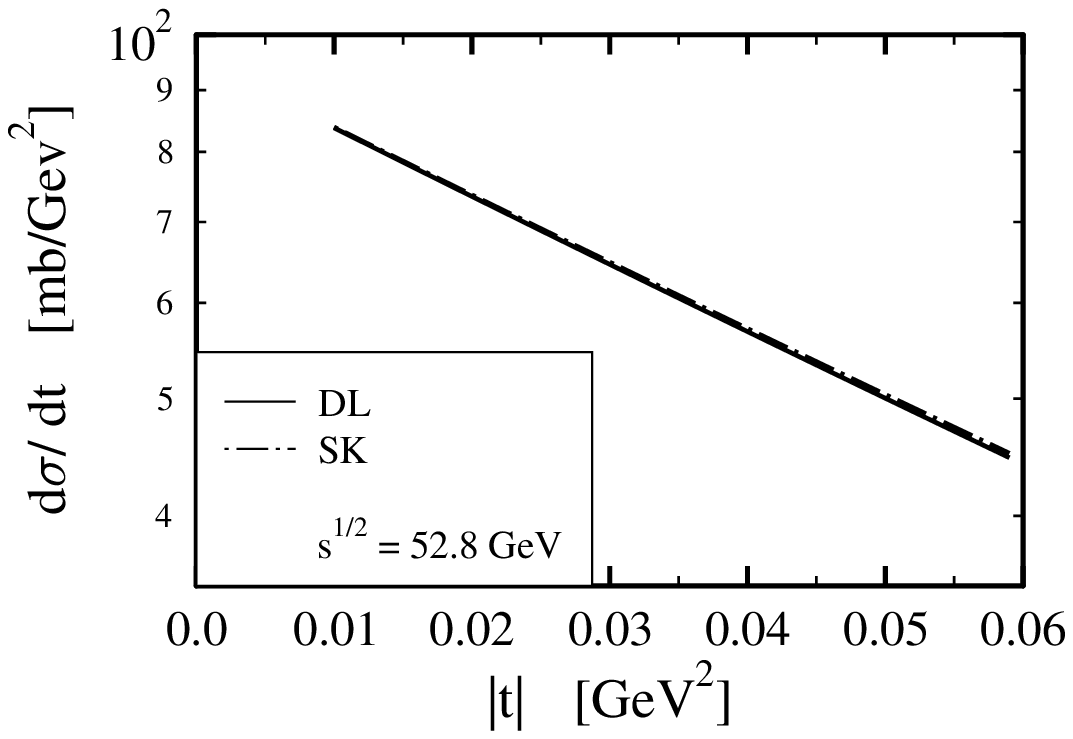}}
   \par
   \caption{Differential cross section for elastic proton-proton
             scattering for $ \protect{\sqrt{s}} = 52.8$ GeV}
   \label{el1}
\end{figure}

\begin{figure}[p]
   \par
   \centerline{\psfig{figure=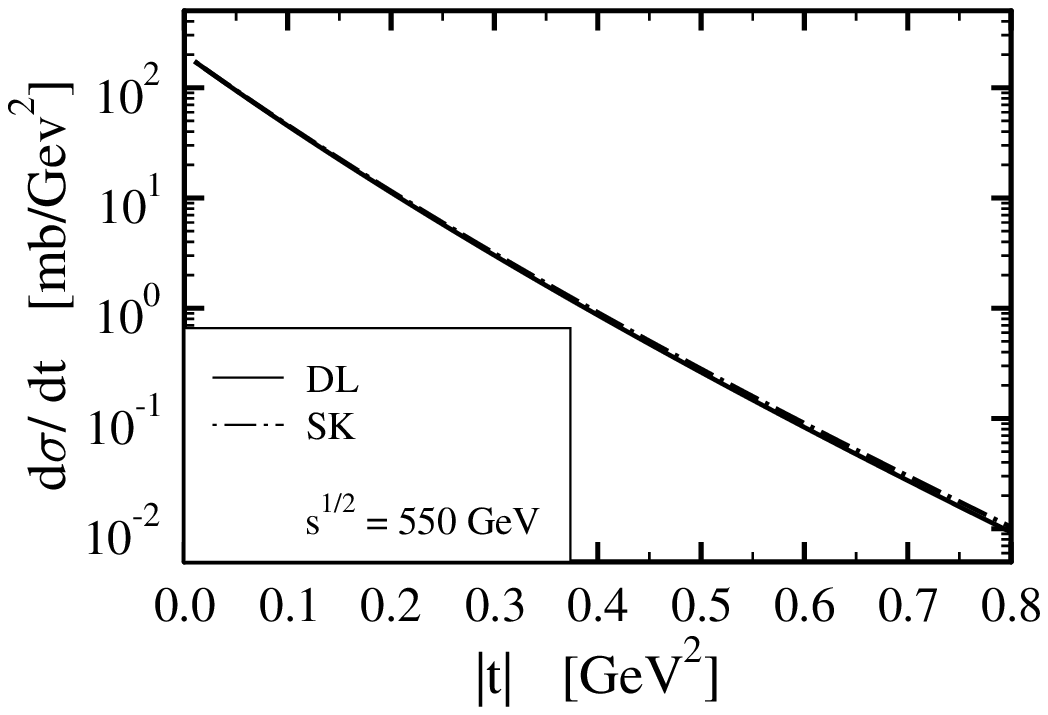}}
   \par
  \caption{Differential cross section for elastic proton-proton
             scattering for $\protect{\sqrt{s}}=550$ GeV}
  \label{el2}
\end{figure}

Another touchstone for any parametrisation of the pomeron are
the cross sections
for elastic proton-proton scattering \cite{l4}.
In figure \ref{el1} the results for the
DL-pomeron and our modified form
are compared for a cms energy $\sqrt{s}=52.8$ GeV
and in figure \ref{el2} the same quantity is plotted for
$\sqrt{s} = 550 \, {\rm GeV}$.
For small values of
momentum transfer $|t|$ the two curves agree very well and the differences
for larger $|t|$ are about one percent.

The good agreement for small values of $|t|$  can be traced
back to the Gordon
identity (\ref{gordon}). For high energy and small momentum transfer the
components of the pomeron's four momentum
are small compared to the energy of the protons.
(The picture of the pomeron is only applicable for
$\omega_{\sPP} < 0.1 \cdot E_{\rm N}$.)
Therefore the
coupling is dominated by the first term, involving the four momenta of
incoming and outgoing proton:
\be
  2 m_N \,
  \gamma_{\alpha} \; = \;
     \left( p + p^{\prime}\right)_{\alpha}
      \, + \, \sigma_{\alpha\alpha^{\prime}}
        \PP^{\alpha^{\prime}}
   \; \approx \; \left( {\it p} + {\it p}^{\prime}\right)_{\alpha} \; .
\ee

In summary we conclude that both the total and elastic cross
sections for nucleon-nucleon scattering
are described equally well by both pomeron vertices.

\section{Diffractive meson production in nucleon-nucleon collisions}

\begin{figure}[p]
    \par
    \centerline{\psfig{figure=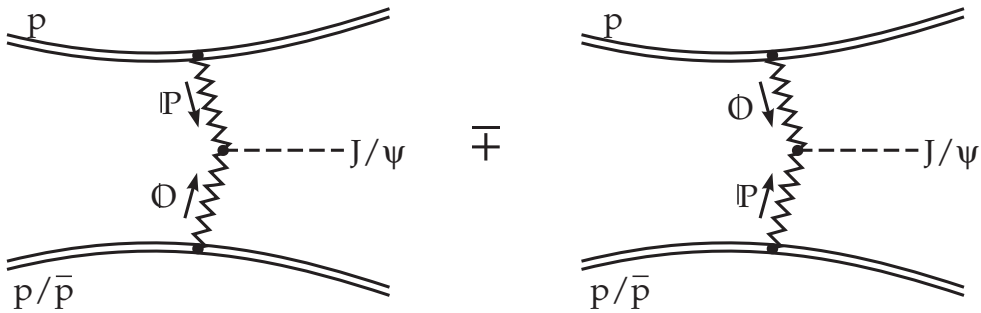,width=13cm}}
    \par
   \caption{Double diffractive $\JP$ production in NN collisions}
  \label{jpsi}
\end{figure}

Another process involving the pomeron is double
diffractive production of $\JP$ mesons in nucleon-nucleon collisions.
In this process the dominant contribution stems from
pomeron-odderon fusion \cite{jpsi1}.
The relevant Feynman diagrams are shown in figure \ref{jpsi}, and
the reader should keep in mind that each diagram represents
two contributions for the different directions of momentum flow
at the $\PP$-$\OO$-$\JP$-vertex. Another point we want to emphasize
is the relative minus sign for $pp$ scattering, while the
two graphs have to be added for $p\bar{p}$.

The odderon is essentially taken to be the $C=-1$ counterpart of the pomeron
\cite{oddi}.
It's coupling to quarks is assumed to be the same as for
the pomeron, namely $\gamma_\alpha$ in the traditional picture and
$\Gamma_\alpha$ for the new coupling, respectively.

The odderon coupling strength is written as $c_{0}\beta$ instead of $\beta$,
with $c_{0} \approx 0.05$ \cite{oddi}.
For it's trajectory we assume a linear form
\be
 \label{odu}
 \alpha_{\sOO} \; = \;
    1 + 0.1 \,{\rm GeV}^{-2} \cdot t_{\sOO} \, .
\ee
The effects of varying the parameters in (\ref{odu}) are small
\cite{jpsi1}.

Again the resulting cross sections are multiplied by a factor
$N$ to take roughly into account the small size of the
$\JP$ meson. For double diffractive processes this factor
is taken to be $N=0.01$.

\begin{figure}[p]
   \par
   \centerline{\psfig{figure=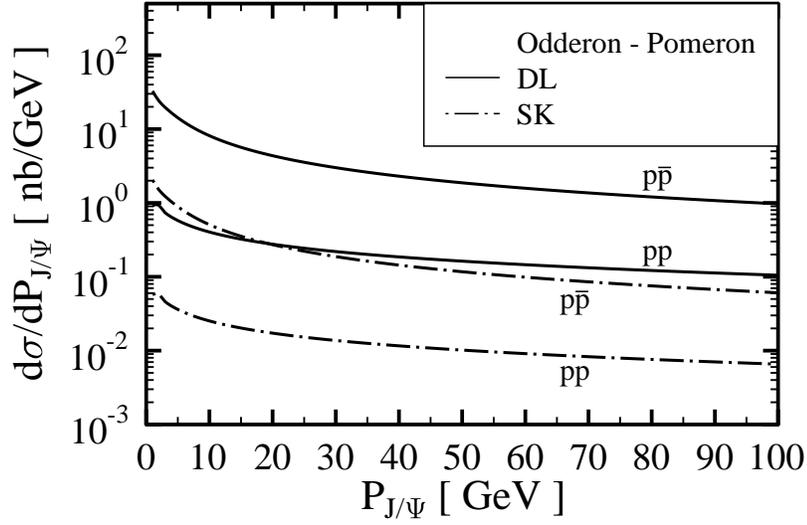}}
   \par
   \caption{Differential cross sections for double diffractive
            $\JP$ production at $\protect{\sqrt{s}} = 2$ TeV}
  \label{sigjpsi}
\end{figure}

In figure \ref{sigjpsi} we show the differential cross sections for
$\JP$ production for $\sqrt{s} = 2$ TeV, which corresponds to
the energies of $1$ TeV for each nucleon. The solid curves
show the results for a pure $\gamma$-coupling lying in the region of
nb. For $p\bar{p}$ scattering the cross section is about one order
of magnitude smaller, since we get destructive interference between the
two graphs of figure \ref{jpsi}.

The signal for the odderon, namely the large ratio between the $pp$
and $p\bar p$ cross section, stays the same
for the new vertex, but the overall size is reduced strongly, making
experiments difficult.

If $c_{0}$ were known (which it is not \cite{c0}) the measurement
of the $\JP$ cross section would be a good probe to decide
between the two pomeron models.

There is, however, another process for $\JP$ production which
offers some possibilities
to look at the pomeron's coupling.
It is the indirect $\JP$ production
via the pomeron-pomeron fusion into one of the $\chi$ mesons.
There exist three of them distinguished by their spin
and each of them can decay into a $\JP$ and a photon \cite{ecki}.
Since the pomeron couples in the same way to $p$ and $\bar{p}$, the
cross sections for diffractive $\chi$ production are the same
for $pp$ and $p\bar p$ scattering.

\begin{figure}[p]
   \par
   \centerline{\psfig{figure=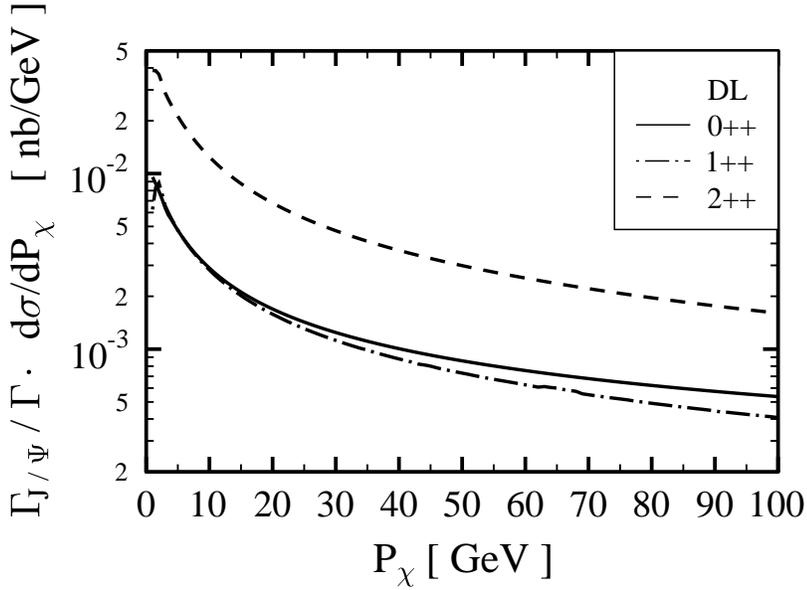}}
   \par
   \caption{Background of $\chi$ mesons obtained with
            $\gamma$-coupling}
  \label{chi1}
\end{figure}

For pure $\gamma$-coupling the background of these processes can be
neglected. The relevant cross sections multiplied by the branching
ratios
\be
 \label{brs}
  \frac{\ds \Gamma_{\JP}}{\ds\Gamma}(0^{++}) \; \approx \; 6.6\cdot 10^{-3}
  \, , \quad
  \frac{\ds \Gamma_{\JP}}{\ds\Gamma}(1^{++}) \; \approx \; 0.273
  \, , \quad
  \frac{\ds \Gamma_{\JP}}{\ds\Gamma}(2^{++}) \; \approx \; 0.135
\ee
are much smaller than $0.1$ nb/GeV, cf.~figure \ref{chi1}.

\begin{figure}[p]
   \par
   \centerline{\psfig{figure=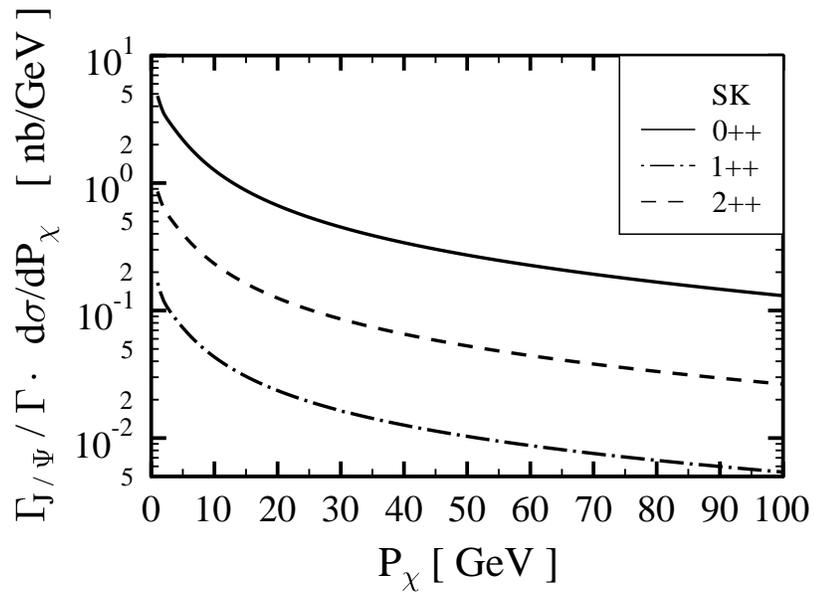}}
   \par
   \caption{Background of $\chi$ mesons obtained with
            the modified coupling}
  \label{chi2}
\end{figure}

For the new vertex the situation is completely
different, now this "background" is contributing
more than
$\PP$-$\OO$-fusion, figure \ref{chi2}.

\begin{figure}[p]
   \par
   \centerline{\psfig{figure=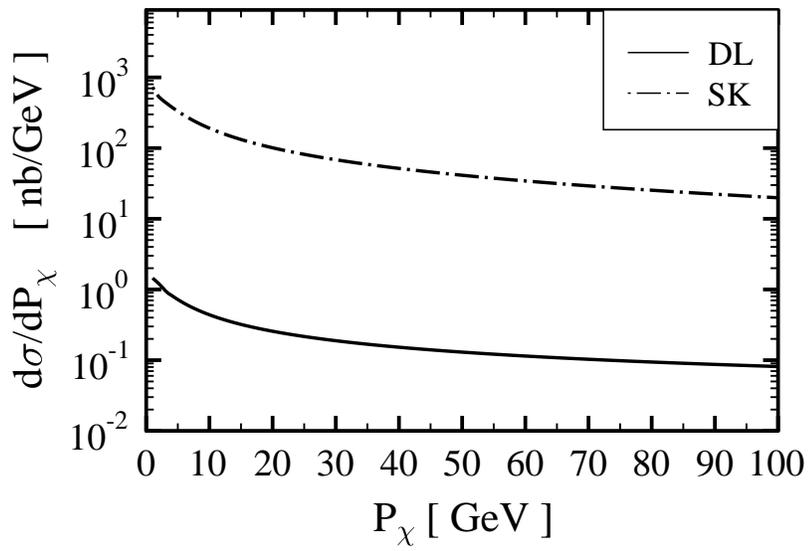}}
   \par
   \caption{Differential cross section for $\chi$ mesons with
            the different couplings}
  \label{chi3}
\end{figure}

The $\JP$ from $\chi$ decay into $\JP + \gamma$ can easily be
discriminated by detection of the photon.
Since the cross section with the $\Gamma$--vertex
is three orders of magnitude larger than the cross section calculated
within the traditional picture
it should be possible
to distinguish between the two pictures, see figure \ref{chi3}.

One way or another, diffractive processes in nucleon-nucleon collisions
give us a new tool to investigate the pomeron in some detail.
Even if the direct $\JP$ production is not accessible to present
experiments, the $\chi$ production could give interesting informations
about the pomeron's coupling.

\section{Exclusive $\rho$ production in DIS}

\begin{figure}[p]
   \par
   \centerline{\psfig{figure=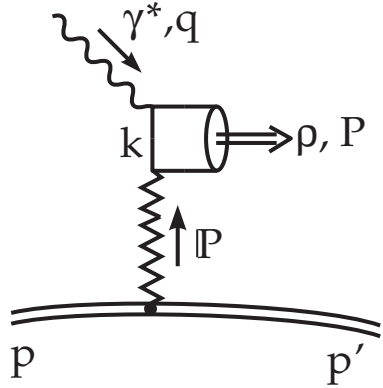,width=5cm}}
   \par
   \caption{Exclusive $\rho$ production in DIS}
  \label{rhotot1}
\end{figure}

Another process where the concept of the pomeron has been used with success
is the exclusive production of $\rho$ mesons in DIS \cite{lrho}.
The relevant graph is similar to that for $\Phi$ production at HERA, but
we consider the scattering of a virtual photon off a proton,
figure \ref{rhotot1}.
The cross section for exclucive $\rho$ production have
been measured by EMC for several
values of
$W^{2} = \left(q + p\right)^{2}$,
and
$ Q^{2} = -q^{2}$, \cite{emc}.
Since the results do not depend strongly on the value of
$W^{2}$, we take for definiteness $W^{2} = 81 \,
{\rm GeV}^{2}$.

The normalization of our cross section is given by the following
expression
\be
 \label{normrho}
 \ba{rcl}
  d\sigma^{(\lambda)} & = & {\ds \frac{4\pi^{2}\alpha}{K}}
                  \, {\ds \frac{1}{4\pi m_{p}}}
                  \, (2\pi)^{4} \delta^{4}\left( q + p - p^{\prime} - P\right)
                  \, (3\beta)^{2} \, F_{N}^{2}(t)
                  \, (\alpha^{\prime} W^{2})^{2\alpha_{\sPP} -2}  \\
             &&  \quad  \sum\limits_{\varepsilon}
                  \varepsilon^{\mu}\varepsilon^{\nu}
                  \, h^{\alpha\beta} \, T^{(\lambda)}_{\mu\alpha}
                                  T^{(\lambda) *}_{\nu\beta} \,
                  {\ds \frac{d^{3}p^{\prime} d^{3}P}
                       {(2\pi)^{6} 4 E_{p}^{\prime} E_{\rho}}}
 \ea
\ee
with $\lambda = 0$ for longitudinally and $\lambda = \pm 1$
for transversely polarised $\rho$ mesons.
The normalization constant $3A^{2}$ can be calculated in analogy
to (\ref{psi0}).
The flux factor $K$ is chosen in the usual way
$K  = (W^{2} - m_{p}^{2}) /  2m_{p}$.
As before the amplitude $T_{\mu\alpha}$ sums up coherently both
directions of momentum flow.
The normalization constant in the pomeron's propagator is fixed
as
$\alpha^{\prime} = 0.25 \,{\rm GeV}^{-2}$
\cite{pommi}.

\begin{figure}[p]
   \par
   \centerline{\psfig{figure=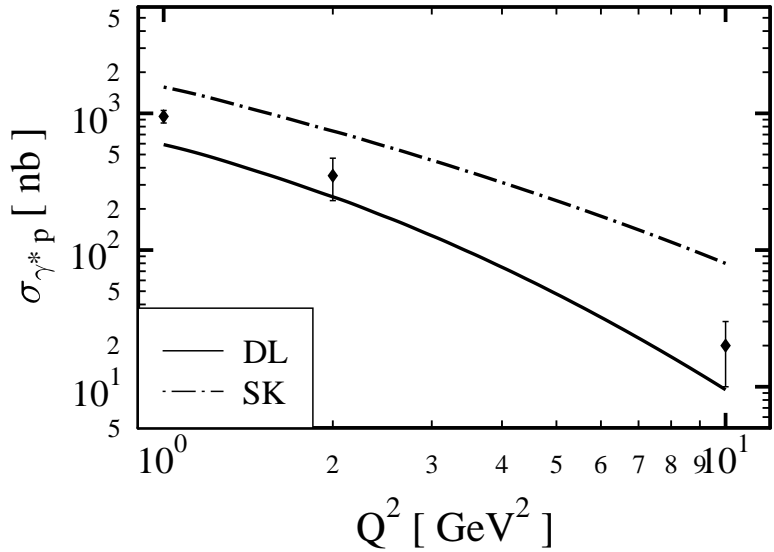}}
   \par
   \caption{Cross section for exclusive $\rho$ production in DIS}
  \label{rhotot2}
\end{figure}

With this expression, keeping the "charge" $\beta$ fixed for all
values of $Q^{2}$, we get the results shown in figure \ref{rhotot2}.
The squares and the error bars mark approximately
the range of the experimental
data for several values of $W^{2}$ and muon energy for
$Q^{2} = 1,2,10 \, {\rm GeV^{2}}$, respectively \cite{emc}.

The cross section calculated with the traditional $\gamma$--coupling (DL)
contains an additional form factor
\be
  \label{lfak}
   {\ds \frac{\mu_{0}^{2}}{\mu_{0}^{2} - k^{2}}} \, ,
   \quad \mu_{0} \approx 1.2 \,{\rm GeV} \, ,
\ee
suppressing the coupling of the pomeron to the quark in the loop.
This quark of momentum $k$, figure \ref{rhotot1},
is far off shell and (\ref{lfak}) was introduced to
get the correct $Q^{2}$ dependence of the cross section \cite{lrho}.
The quantitiy $\mu_{0}$ is fixed by the relation of
pomeron exchange and the proton's structure function
$F_{2}(x)$ for small $x$.

With these ingredients
the shape of the experimental data is reproduced very well,
but the absolute values
differ approximately by a factor
of two.
This is in contrast to the results quoted in \cite{lrho},
which are in good agreement with experiment. To make
a comparison possible we give our normalization of the cross section
(\ref{normrho}).

The cross section calculated with the $\Gamma$-vertex (SK)
is about two times larger than the experimental data.
Bearing in mind that the applicability of our model is
questionable due to the large
$p_T$ of the $\rho$ and the smallness of the up and
down quark masses this agreement is quite encouraging.
The shape of the results is
compatible with experiment, and we want
to emphasize that this result is calculated without any additional
form factor
(\ref{lfak}).

\begin{figure}[p]
   \par
   \centerline{\psfig{figure=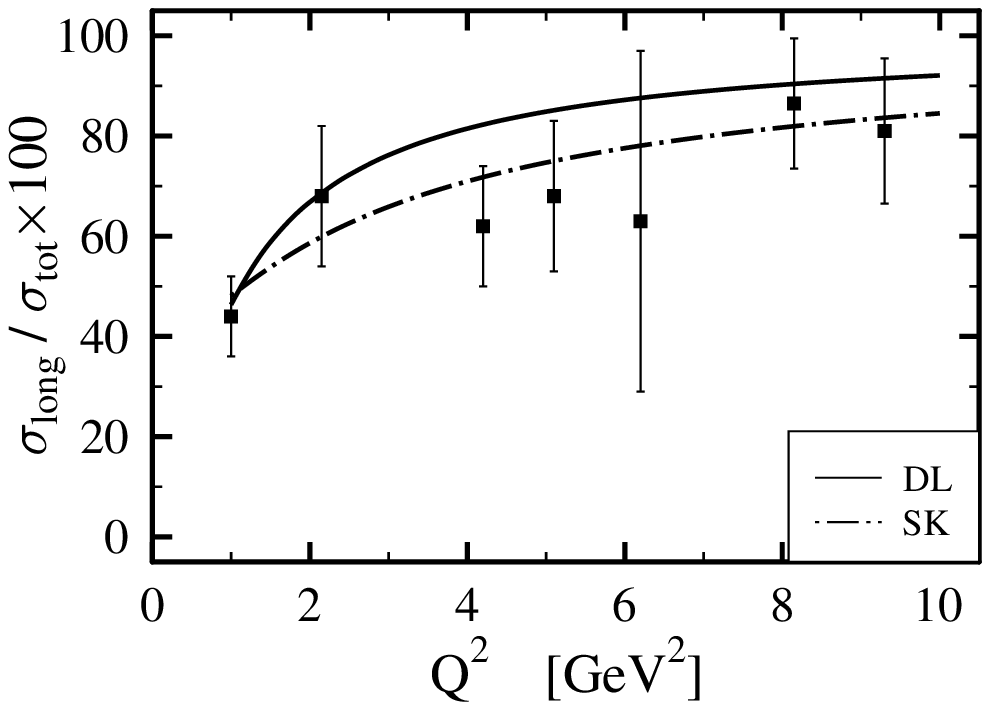}}
   \par
   \caption{Fraction of longitudinally produced $\rho$ mesons}
  \label{rhotot3}
\end{figure}

Another quantity measured by EMC is the fraction of longitudinally
polarised $\rho$ mesons defined by
\be
  \frac{\sigma_{long}}{\sigma_{tot}} \; = \;
      \frac{\sigma^{(\lambda = 0)}}
             {\sum\limits_{\lambda = 0,\pm1}\sigma^{(\lambda)}} \, ,
\ee
The results for both couplings are shown in figure \ref{rhotot3}.
With the traditional $\gamma$--coupling the polarization of the
$\rho$--meson is described in good agreement.

With the new coupling the experimental data are
described within error bars. This property depends
crucially on the concrete definition of (\ref{wawa}),
respectively (\ref{jogi}). This representation apparently shows
a suppression of the momentum dependent term
for higher momentum transfers, and this suppression
is responsible for the resulting
$Q^2$--dependence of the polarization.

Finally we want to stress that the new vertex describes correctly
diffractive nucleon--nucleon
scattering, $\Phi$ production and $\rho$--production without any
further assumption and that the combination in (\ref{wawa}) is unique
in doing so.
This effective quark--pomeron coupling should guide QCD based
investigations of the pomeron.

\section{Summary}

The investigation of the pomeron is currently of great interest.
Attempts to derive its properties from QCD in a rigorous way are faced
with enormous technical difficulties \cite{hans}. Therefore experiments
feasible, e.g.~at HERA, can only be sensibly compared
with results from phenomenological models like
the Donnachie--Landshoff pomeron. We proposed a modified pomeron
model which reproduces all the successes of the DL--pomeron.
Our result is unique, and
avoids a number of problems
arising due to the violation of gauge invariance.
We have shown that diffractive meson production is
very sensitive in discriminating between the two models.
Additionally our vertex leads to spin--effects which, although small,
might be observable at HERA.
If the phenomenologically correct
vertex structure were determined by such experiments this could
guide the effort of deriving the pomeron from QCD.

\vspace{0.3cm}

\begin{center}
   {\bf Acknowledgement}
\end{center}
\vspace{-0.3cm}
This work was supported by the Deutsche Forschungsgemeinschaft
(G.~Hess program). We thank P.V.~Landshoff for his help
and very useful discussions.

\newpage

\clearpage

\newpage

\begin{thebibliography}{88}


  \bibitem[1] {hera}
    G.~Wolf, DESY preprint 94--022

  \bibitem[2] {f2}
    I.~Abt et al., Nucl..~Phys.~{\bf B 407} (1993) 515 \\
    I.~Abt et al., Phys.~Lett.~{\bf B 321} (1994) 161

  \bibitem[3] {hans}
    H.~Lotter, J.~Bartels,
    Phys.Lett.~{\bf B 309} (1993) 400 \\
    J.R.~Forshaw, P.N.~Harriman, P.J.~Sutton,
    Nucl.Phys.~{\bf B 416} (1994), 739

  \bibitem[4] {jpsi1}
    A.~Sch\"afer, L.~Mankiewicz, O.~Nachtmann,
    Phys.Lett.~{\bf B 272} (1991) 419

  \bibitem[5] {ecki}
    E.~Stein, A.~Sch\"afer,
    Phys.Lett.~{\bf B 300} (1993) 400

  \bibitem[6] {kuhn}
    J.H.~K\"uhn, J.~Kaplan, E.G.O.~Safiani,
    Nucl.Phys.~{\bf B 157} (1979) 125

  \bibitem[7] {berger}
    E.L.~Berger, D.~Jones, Phys.Rew.~{\bf D 23} (1981) 1521

  \bibitem[8] {higgs}
    A.~Bialas, P.V.~Landshoff, Phys.~Lett.~{\bf B 256} (1991) 540

  \bibitem[9] {pommi}
    A.Donnachie, P.V.~Landshoff, Phys.Lett.~{\bf B 123} (1983) 345;
    Nucl.Phys.~{\bf B 231} (1984) 189; Nucl.Phys.~{\bf B 244} (1984) 322;
    Nucl.Phys.~{\bf B 267} (1986) 690

  \bibitem[10] {l1}
    P.V.~Landshoff, University of Cambridge preprint DAMTP/88-27

  \bibitem[11] {bjorken}
    C.~Itzykson, J.-B-~Zuber, {\it Quantum Field Theory},
     New York 1980

  \bibitem[12] {jpex}
    T.~Ahmed et al., Phys.Lett.~{\bf B 338} (1994) 507

  \bibitem[13] {parton}
    P.V.~Landshoff, J.C.~Polkinghorne, Phys.Rep.~{\bf C 5} (1972) 1

  \bibitem[14] {l4}
    A.~Donnachie, P.V.~Landshoff, Nucl.Phys.~{\bf B 231} (1984) 189

  \bibitem[15] {oddi}
    A.~Donnachie, P.V.~Landshoff,
    University of Cambridge preprint DAMTP/90-17

  \bibitem[16] {c0}
    R.J.M.~Covolan,P.~Desgrolard, M.~Giffon, L.L.~Jenkovszky,
    E.~Predazzi, Z.Phys.~{\bf C 58} (1993) 109

  \bibitem[17] {lrho}
    A.Donnachie, P.V.~Landshoff,
    Phys.Lett.~{\bf B 185} (1987) 403

  \bibitem[18] {emc}
    J.J.~Aubert et al., Phys.Lett.~{\bf B 161} (1985) 203



\end{thebibliography}
\end{document}